# AN ALKALI-VAPOR CELL WITH METAL COATED WINDOWS FOR EFFICIENT APPLICATION OF AN ELECTRIC FIELD

D. Sarkisyan and A.S. Sarkisyan

Institute for Physical Research, Armenian Academy of Sciences,

Ashtarak2, 378410, Armenia

J. Guéna, M. Lintz[a] and M.-A. Bouchiat

Laboratoire Kastler Brossel[b] and Département de Physique de l'Ecole Normale Supérieure[c],

24 Rue Lhomond, 75231 Paris cedex 05, France

**ABSTRACT**

**We describe the implementation of a cylindrical T-shaped alkali-vapor cell for laser spectroscopy in the presence of a longitudinal electric field. The two windows are used as two electrodes of the high-voltage assembly, which is made possible by a metallic coating which entirely covers the inner and outer sides of the windows except for a central area to let the laser beams in and out of the cell. This allows very efficient application of the electric field, up to 2 kV/cm in a rather dense superheated vapor, even when significant photoemission takes place at the windows during pulsed laser irradiation. The body of the cell is made of sapphire or alumina ceramic to prevent large currents resulting from surface conduction observed in**

**cesiated glass cells. The technique used to attach the monocrystalline sapphire windows to the cell body causes minimal stress birefringence in the windows. In addition, reflection losses at the windows can be made very small. The vapor cell operates with no buffer gas and has no magnetic part. The use of this kind of cell has resulted in an improvement of the signal-to-noise ratio in the measurement of Parity Violation in cesium vapor underway at ENS, Paris. The technique can be applied to other situations where a brazed assembly would give rise to unacceptably large birefringence in the windows.**

## I. INTRODUCTION

Glass cells have been widely used for spectroscopy of alkalis, offering the possibility of an atomic number density considerably higher than with atomic beams or trapped atoms. In addition one can perform laser spectroscopy of an alkali vapor in the presence of a transverse electric field using internal electrodes and glass-metal electrical feedthroughs. The first measurements of the Cs 6S-7S parity violating electric dipole transition amplitude due to Atomic Parity Violation (APV) were made by observing the fluorescence of a cesium vapor excited by a cw 539.5 nm laser beam in the presence of a transverse dc electric field of 200 V/cm[1]. More recently, in order to improve the measurement precision, a new APV experimental design based on a pulsed pump-probe scheme, with a *longitudinal* electric field operated in a pulsed regime, has been implemented[2]. In this experiment, however, the operating conditions were more demanding (2kV/cm electric field, intense excitation pulses of 1.8 mJ in 15ns) and glass cells exhibited several significant drawbacks, due to the interaction of the glass surface with the chemically aggressive alkali vapor.

First, although they are considered to be good insulators, silica and silicate glasses exhibit a considerable surface conductivity when alkali vapor is present. Indeed the square resistance can be as low as 10 k$\Omega$ for the inner wall of a synthetic silica cell in the presence of a $10^{13}$ at/cm$^3$ cesium vapor[3]. This can give rise to large conduction currents when high electric fields are applied. Second, the optical transparency of glass is lost if the glass surface is exposed to high temperature alkali vapor or, in the longitudinal electric field configuration, to electron/ion bombardment at the windows, as a result of photoionization of cesium dimers by the intense pump laser pulses[4].

Cells with sapphire windows and a sapphire body attached together by a high temperature, electrically insulating binder made of a melted powder of metal oxides and carbonates[5,6] provide a significant improvement in these respects. In sapphire cells, measured surface conductivities are several orders of magnitude lower than in glass cells[3]. This allows one to apply an electric field *inside* a cesium cell using *external* electrodes[4]. Furthermore sapphire is known to be more resistant to radiation damage than silica[7,8]. Superheated cesium vapor has no effect on the optical transparency of sapphire up to 1000 °C [5, 9], nor does electron/ion bombardment[4]. Sapphire is also often preferred to glass for its wider transparency spectral range.

Although sapphire cells show excellent insulating properties, we have encountered problems linked with space charge effects in the presence of a longitudinal electric field, when shining an intense pulsed laser beam through the windows[10]. The reason for this space charge was shown to be a considerable multiplication of the initial charge photoemitted at the "cathode" window (close to the negatively biased electrode). Grooving the inside wall of the cell turned out to be very efficient against the multiplication of the photoelectron charge, which confirmed that the multiplication was due to secondary electron emission when the accelerated photoelectrons hit the cell wall at grazing incidence. Such a grooved cell made possible the first measurement of APV in cesium vapor using the pump-probe scheme in the longitudinal electric field configuration[11].

If grooving the cell has been efficient in suppressing the *multiplication* of the electronic space charge, however the primary photoemission remains, and is likely to affect, to some extent, the longitudinal electric field map close to the window, though with a much lesser extent than with non-grooved cells. Photo-emission leaves a positive charge at the cathode window, and the accelerated electrons accumulate at the anode window, giving rise to a negative charge. Preventing photoemission in the visible range seems difficult to achieve. Cesium is known to adsorb at the alumina surface[12], and a notable decrease of the electron workfunction has been recorded on alumina exposed to cesium vapor (-2eV shift)[13]. Indeed a 1.4 eV electron workfunction was measured for cesiated sapphire[14], a value smaller than the 2.3 eV photon energy at 539 nm.

To prevent the inner side of the windows from charging during photoemission, the best solution would be a conductive treatment covering the whole window surface. The coating would ensure electrical continuity between the window's inner and outer surfaces, the latter being connected to the voltage generator. This would allow one to supply the current required to maintain the potential at the window, and hence the value of the applied longitudinal electric field in the vapor despite photoemission at the windows. Unfortunately, optically transparent conductive coatings like indium-tin oxide (ITO) do not withstand dense alkali vapor. For this reason, we have chosen to apply a thick metallic coating, completely opaque, *at the periphery* of both sides, leaving an open, uncoated area for the transmission of the laser beams. As opposed to the use of a *semi-transparent* metallic coating, this preserves the possibility of efficiently canceling laser beam reflection at the windows, using temperature tuning of the interference order[15], and accordingly, boosting the transmission up to nearly 100%.

This paper describes the realization and implementation of the cesium cell with metal coated windows that has been used at Ecole Normale Supérieure (ENS) in the 2.7% precision APV experiment recently reported[16], for which the following severe constraints had to be met simultaneously:

    1) The longitudinal electric field should be as close as possible to the ideal, homogeneous field.

    2) Birefringence in the windows should be low: 1 or a few mrad per window.

    3) Geometrical imperfections should be kept low; in particular the windows should be perpendicular to the tube axis to within 1 mrad.

## II. DESIGN AND REALIZATION OF THE CELL

**Attempts at a sapphire-to-niobium brazed assembly**

Before addressing the design and implementation of the metal coated cell, we should mention several attempts at preparing a cell made of 7 insulating sections separated by 6 ring-shaped metal electrodes. The parts were to be assembled by brazing. Since the attack of brazing material in alkali cells has been suspected many times[17], we first exposed a sample of Ag-Cu brazing alloy[18] to a 10 mTorr *unsaturated* cesium vapor at 280°C during three weeks, for a study of the penetration of Cs inside the material. X-ray photoelectron spectroscopy measurements made at CERN, Geneva, showed[19] that Cs could be observed at a depth of 0.6 nm and, as a trace element, up to 12 nm, but not deeper. Energy-dispersive X-ray microscopy does not reveal any difference between the surface of the exposed sample and the surface of a reference sample. This suggests that the suspected corrosion problems may be due to a contact with the *liquid* alkali rather than with the vapor, and could be avoided by preventing the liquid from reaching the brazed parts.

However, as the windows were brazed on the niobium end-electrodes (Fig. 1), one obvious question was how much stress birefringence would be added by the brazing process. Although the thermal expansion coefficient of niobium is very close to that of sapphire, the difference is not negligible. Brazing is achieved by heating together the parts and the brazing material at a temperature of ≈800 °C (for Ag-Cu 72%-28% brazing material) or above. Since the experiment takes place at a much lower temperature, of about 250 °C, any difference in the expansion curves can give rise to stresses in the window, and hence potentially birefringence. Even a birefringence phase difference as low as 0.01 rad is too large for the APV experiment[16], unless it can be compensated. Attempts were made with different brazing materials and brazing technologies and designs. Only one of the test pieces had a birefringence ≤ 0.01 rad. But this result could not be reproduced and all the other tests gave values > 0.01 rad, up to 0.1 rad. Birefringence was not only large, but also inhomogeneous in size while the direction of the optical axes was variable, which definitely prevented the use of birefringence compensators.

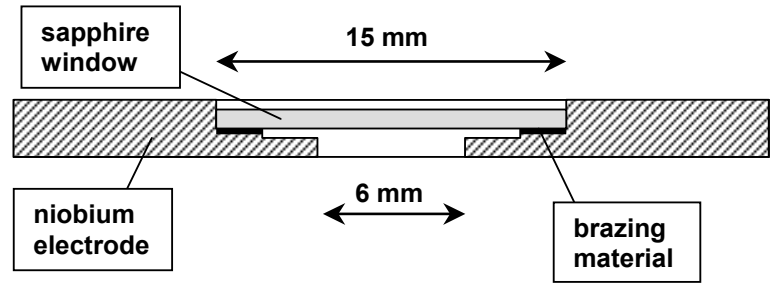

*Fig 1: Schematics of the attempted brazed window assembly. Actually, different types/designs of brazing were tested, in order to minimize the stress to the window, but none was sufficiently reproducible, as regards our birefringence requirements, to be finally adopted.*

**The choice of the sapphire-to-sapphire assembly**

As opposed to the brazed assemblies mentioned above, all sapphire cells assembled as described in [5] have shown values of the birefringence of the order of 1 or a few mrad, reasonably uniform so that it could be reduced using a birefringence compensator. For this reason we have considered the design of Fig. 2, where the metallic coating allows electric continuity between the inside and the outside of the cell, and, from there, to the high voltage (HV) generator, using a knitted Monel ring (as shown by Fig. 2 in [10]). Thus the photoelectrons that reach the metal-coated part of the anode window can be extracted from the cell, and the positive charge left at the cathode window is more efficiently compensated, allowing better control of the applied electric field.

The body of the cell ($L$ = 83 mm; $\varnothing_{out}$ = 13 mm; $\varnothing_{in}$ = 10 mm) is a tube obtained from a 99.7% pure alumina ceramic[20] rod. The inner surface of the tube was machined to obtain grooves, so as to reduce secondary electron emission under grazing incidence[10]. A sapphire side arm ($L$ = 30 mm; $\varnothing_{out}$ = 3 mm; $\varnothing_{in}$ = 1 mm) is glued in a hole made at the center of the tube and terminated by a glass reservoir ($L$ = 40 mm; $\varnothing_{out}$ = 5 mm; $\varnothing_{in}$ = 4 mm).

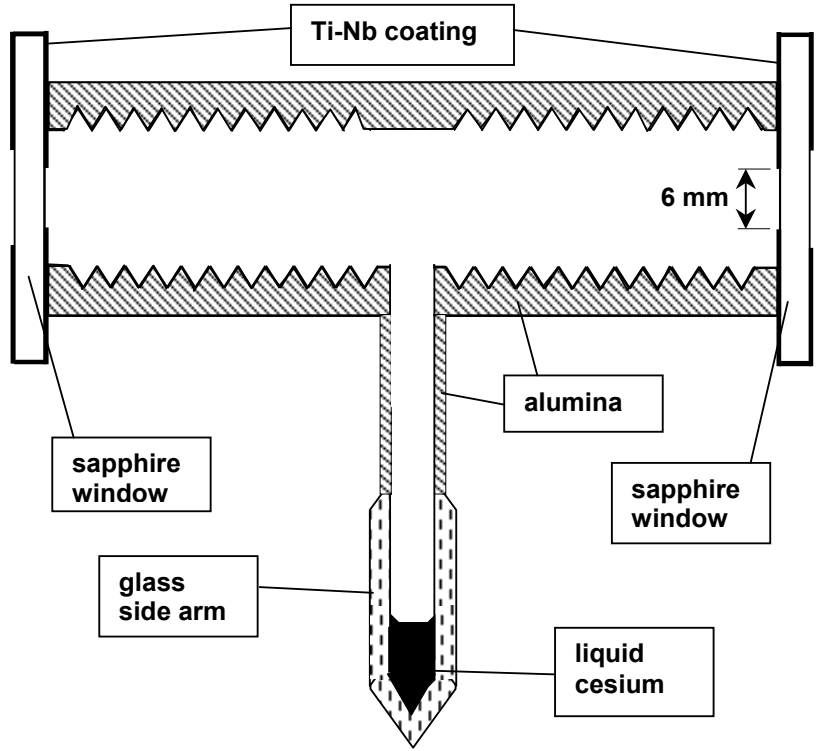

*Fig 2: Drawing of the implemented sapphire/alumina cell with two metal-coated windows. The thick dark line around the windows represents the metal coating, actually a few microns thick.*

**Preparation of the windows**

The 0.5 mm thick windows[21] are made from highly parallel sapphire plates (parallelism better than 10μrad, allowing efficient cancellation of the reflection to better than $2 \times 10^{-3}$ of the incident laser beam power[15]), with (0001) orientation to better than 0.25°. The high quality polishing by the manufacturer, together with an annealing at 1250 °C for 2 hours ensures a clean surface with reproducible stoechiometry and physical properties. In particular, annealing gives rise to the "reconstruction" of the surface at the atomic scale, which gives rise to flat terraces, several hundred nanometers long, separated by monoatomic steps, as evidenced by atomic probe microscopy[22,23] or by laser beam diffraction[24]. This reduces the probability of cesium adsorption at corners, edges...[13]

The sapphire plates received a titanium (2μm thick) + niobium (2μm thick) coating obtained by RF sputtering[25]. Except for an open central area of 6 mm diameter (Fig. 2), the coating covers the whole surface of the window, in particular the edge. As first layer we have chosen titanium which strongly interacts with alumina[26]. Niobium has been chosen as the material for the cover layer because of its excellent getter properties, of its good electrical conductivity, and because its electron emission work function is higher than that of titanium[27] in the 200-270 °C temperature range used in the APV experiment at ENS. Both metals are non-magnetic.

Since RF sputtering is obtained by a discharge in an argon atmosphere, it is likely that the metallic coating has a non-negligible Ar content. To prevent or reduce possible Ar outgassing during spectroscopic operation of the cell at 250 °C, prior to assembling them with the rest of the cell the coated windows are outgassed in vacuum at 300°C for several hours.

Metal deposition on a substrate is known to induce stress in the substrate. Birefringence was measured before and after Ti-Nb deposition. In fact, no significant change was observed: in the open area the birefringence phase difference remained close to 1 mrad or lower before gluing.

**Assembling the cell**

Before gluing[28] the sapphire plates, a number (16) of small open circle areas of 1mm diameter were prepared by removing the Ti-Nb coating. The idea is the following: since sapphire surface can be glued reliably to an alumina ceramic (or monocristalline sapphire) tube using the technique presented in [5], the Ti-Nb coating is removed in some places to ensure direct contact of sapphire-plate to the ceramic tube for the purposes of gluing. This appears as a reasonable precaution although niobium seems reasonably wetted by the binding material. Gluing of the Ti-Nb coated sapphire windows to the alumina tube has been realized by applying the binder material at the junction between the tube and window, and heating, in a vacuum chamber[5], at 1200°C.

Before we fill it with Cs the empty cell is sealed off in order to check how well it holds vacuum. A small leak might arise from a micro-crack in the alumina ceramic body. The quality of the vacuum in the cell disconnected from the pump can be tested by using a simple discharge and observing visually (in a dark room) the weak fluorescence. Several days without a change in the fluorescence from the residual gas are enough to ascertain the absence of a leak. From our point of view this is one of the sensitive ways to detect any small leakage when no residual gas analyzer is available in the laboratory. A micro-crack, if subsequently located, can be filled by applying a thin layer of the binding material, and proceeding once again to the heating under vacuum.

The cell was thoroughly outgassed at the temperature of 270°C (at higher temperature Nb oxidizes in air) for 10 hours before filling it with cesium by inserting a metallic droplet into the side arm. During the APV measurements, a special oven is used to stabilize independently the temperatures of the entrance and exit windows ($T_{in}$ and $T_{out}$, respectively, to within 0.1 °C for reflection cancellation, both in the range 230-260 °C) and of the side arm ($T_{sa} < T_{in}, T_{out}$), the latter controlling the cesium vapor density ($N_{Cs} = 10^{14}$ at.cm$^{-3}$ for $T_{sa} \approx 140°C$)

## III. CHARACTERIZATION

The angle between the windows normal and the tube axis is found to be <1 mrad, as with non-coated windows. This fulfills requirement #3 above (see section I), which would have been hard to obtain by brazing together 17 different metal and insulator pieces (see section 2-).

The birefringence of the whole cell (two windows), as measured using a helium-neon laser, ranges between 4 and 10 mrad, with a minimum at the center of the uncoated area, and the axes have a quite uniform direction. The APV setup at ENS provides for *in-situ* measurements using atomic signals proportional to the birefringence of the *entrance* window. The measured birefringence of a few milliradians, could satisfactorily be compensated. This fulfills requirement #2 (section I).

Measuring the longitudinal electric field inside the cell involves measuring the linear dichroism of the probe laser beam resonant for the 7S-6P$_{3/2}$ transition after linearly polarized excitation of the forbidden 6S-7S transition. The linear dichroism signal $\propto (\beta^2 E^2 + M_1^2)$ has two contributions from the 7S alignment: one associated with the electric field ($\beta$ is the 6S-7S vector polarizability) and a second due to the forbidden magnetic dipole $M_1$ transition . In the presence of an electric field of 1 or 2 kV/cm, the magnetic dipole contribution is negligible ($< 10^{-3}$) as compared to the Stark-induced one. With the electric field turned off only the $M_1^2$ term remains. In this way we determine the field E in terms of the ratio $M_1/\beta$ known to better than 1%[29,30]. Indeed this calibration of the longitudinal electric field applied to the atoms is a necessary step for exploiting the measurements of the APV linear dichroism[16].

If we ignore photoemission, we can calculate the electric field numerically for the actual geometry of the cell and HV setup[4]. For an expected field of 1660V/cm, the linear dichroism data recorded in the APV experimental conditions indicate that the average longitudinal electric field experienced by the atoms is 1626 ± 16 V/cm, that is 98 ±1 % of the calculated value. With an exactly similar cell, except for the absence of the metal coatings at the windows, this value drops to 92 ± 1 % under the same conditions. As was expected, the electric field in the metal-coated cell is much less affected by the photoemitted charges, an improvement which we attribute to partial elimination and compensation of the charges. Symmetry of the applied field after sign reversal remains excellent (asymmetry of order $10^{-3}$), another critical parameter for APV experiments.

Using another atomic signal (a Faraday effect induced by a longitudinal magnetic field gradient $B_z(z) \propto z$, to be compared with the one induced by a uniform field $B_z(z) = Const.$) one can measure the $+z/-z$ asymmetry of the applied electric field. In [10] we have demonstrated a significant inhomogeneity of $E_z^2$ corresponding to an accumulation of electron charges at the anode window. As explained in sect.1, this asymmetry was substantially reduced in grooved cells. We have repeated this kind of measurements in grooved cells *with metal coated windows* and observed that the $+z/-z$ asymmetry is further reduced by a factor $\geq 3$, giving confidence in the beneficial role of the metal coating.

This kind of cell is the one that provided the best signal/noise in the APV experiment underway at ENS, Paris[16].

## IV. OUTLOOK: USE OF THE METAL COATING FOR A TRANSVERSE FIELD CELL

Finally we would like to address the question of the feasibility of the cell drawn in Fig. 3. Such a cell has been considered[31] for APV measurements in cesium vapor, using a *transverse* electric field, where one could benefit from larger excited-state atom numbers, and hence a larger amplification of the left-right asymmetry to be detected. The quantum shot noise limited precision for an APV measurement in transverse E field could reach the $10^{-3}$ level, and systematic effects would become a crucial issue. The reason for the design of Fig. 3 is that it would allow easy rotation of the direction of the transverse electric field by steps of 45°, so as to suppress some possible sources of systematic effects on the APV measurement[31].

The eight metal-coated longitudinal stripes have to be connected to the outside in order to remove the charges possibly present in the vapor. As in the longitudinal field configuration, the gluing of the windows will offer the possibility of implementing the HV feedthroughs to the metal stripes, just by coating radial sections at the ends of the tube, as illustrated on fig. 3. In this way the transverse E field should be applied inside the cell over the whole length of the cell, with no edge effect at the ends if an outer, dummy extension (without alkali vapor) with the same types of connections is added at each end[32]. As in the longitudinal $\bar{E}$ field case, no troublesome birefringence is expected either. Again, the requirements with birefringence and electrical continuity would be hard to meet using a brazed assembly.

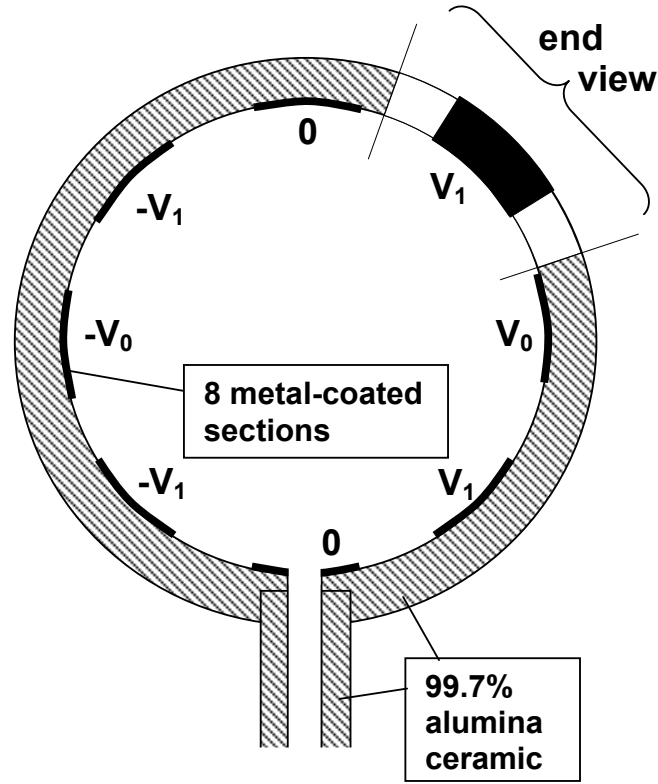

*Fig 3: Transverse section of a cylindrical cell allowing rotation, by steps of 45°, of the transverse applied electric field (by suitable choice of the ratio of $V_0$ and $V_1$, the electric field in the central region of the cell can be made homogeneous). Cutaway view through the middle, except for the upper right section, which is an end view, showing the metal coating of an end section of the cell tube.*

The technique of sapphire-to-sapphire insulator bonding has previously extended the possibilities of atomic spectroscopy to high temperature and high pressure of chemically aggressive vapors[6,8]. It has also provided a new kind of spectroscopic tools, such as for instance sub-micron thick vapor cells giving easy means of access to one-laser sub-Doppler spectroscopy[33]. Extension of this technique with metal-coated windows provides us today with a key-element for spectroscopic studies of a highly forbidden transition such as the 6S-7S Cs transition and precise measurements of the parity violation effect in a longitudinally applied electric field. In future, it should be useful in implementing a cell with application of an electric field, without edge effects, in a different geometry.

## Acknowledgments:

We are indebted to S. Mathot and C. Benvenuti (TS/MME, CERN, Geneva) for help in the sapphire-to-niobium brazing tests.

# REFERENCES

[a] E-mail: lintz@lkb.ens.fr

[b] Laboratoire de l'Université Pierre et Marie Curie et de l'Ecole Normale Supérieure, associé au CNRS (UMR 8552)

[c] Unité FR 684 du CNRS

[1] M.-A. Bouchiat, J. Guéna and L. Pottier, J. Phys. (France), **46** (1985) 1897.

[2] J. Guéna, D. Chauvat, Ph. Jacquier, M. Lintz, M. D. Plimmer and M. A. Bouchiat, Quantum Semiclass. Opt. **10** (1998) 733.

[3] M.-A. Bouchiat, J. Guéna, Ph. Jacquier, M. Lintz and A. V. Papoyan, Appl. Phys. **B68** (1999) 1109.

[4] E. Jahier, J. Guéna, Ph Jacquier, M.Lintz and M.-A. Bouchiat, Eur. Phys. J. **D13** (2001) 221.

[5] D. Sarkisyan and A. Melkonyan, Instrum. Exp. Tech. **32** (1989) 485.

[6] J. A. Neuman, P. Wang and A. Gallagher, Rev. Sci. Instrum. **66** (1995) 3021.

[7] J. Olivier and R. Poirier, Surf. Sci. **105** (1981) 347.

[8] B. Carrière and B. Lang, Surf. Sci. **64** (1977) 209.

[9] J. Schlejen, J. Post, J. Korving and J. P. Woerdman, Rev. Sci. Instrum. **58** (1987) 768.

[10] J. Guéna, E. Jahier, M. Lintz, A. Papoyan, S. Sanguinetti and M.-A. Bouchiat, Appl. Phys. **B75** (2002) 739.

[11] J. Guéna, D. Chauvat, Ph. Jacquier, E. Jahier, M. Lintz, S. Sanguinetti, A. Wasan, M.-A. Bouchiat, A. V. Papoyan and D. Sarkisyan, Phys. Rev. Lett. **90** (2003) 143001.

[12] J. A. Rodriguez, M. Kuhn and J. Hrbek, J. Chem. Phys. **100** (1996) 18240.

[13] M. Brause, D. Ochs, J. Günster, T. Mayer, B. Braun, V. Puchin, W. Maus-Friedrichs and V. Kempter, Surf. Sci. **383** (1997) 216.

[14] A. V. Papoyan, J. Guéna, M. Lintz and M.-A. Bouchiat, Eur. Phys. J. AP **19** (2002) 15.

[15] E. Jahier, J. Guéna, Ph. Jacquier, M. Lintz, A. V. Papoyan and M.-A. Bouchiat, Appl. Phys. **B71** (2000) 561.

[16] J. Guéna, M. Lintz and M.-A. Bouchiat, arXiv:physics/0412017, to appear in Phys. Rev. A.

[17] According to specialists involved in atomic clocks and from litterature: P. Cerez, private communication; M. Baldy, private communication concerning brazing in the presence of liquid cesium; Ref.[9] reports that Cu and Ni O-rings developed leaks in a high temperature reservoir for a sodium vapor cell.

[18] Cusil-ABA active brazing alloy from Wesgo (www.wesgometals.com).

[19] S. Mathot and J. M. Dalin, private communication.

[20] AL23 vacuum-tight ceramic, from Degussa-Hüls.

[21] Meller Optics, 120 Corliss Street, Providence, RI 02904, USA, www.melleroptics.com.

[22] M. Yoshimoto, T. Maeda, T. Ohnishi, H. Koinuma, O. Ishiyama, M. Shinohara, M. Kubo, R. Miura and M. Miyamoto: Appl. Phys. Lett. **67** (1995) 2615.

[23] J. R. Heffelfinger, M. W. Bench and C. B. Carter, Surf. Sci. **370** (1997) L168.

[24] M. Lintz and M.-A. Bouchiat, Surf. Sci. **511** (2002) L319.

[25] Saint Gobain Céramiques Avancées, Rue Marcellin Berthelot, Zone Industrielle, 77550 Moissy-Cramayel, France.

[26] F. S. Ohuchi and M. Kohyama, J. Am. Ceram. Soc. **74** (1991) 1163.

[27] R. G. Wilson, J. Appl. Phys. **37** (1966) 3161 and **37** (1966) 4125.

[28] The term "gluing" will be used although the material used is not properly a glue, but a high temperature binder.

[29] S. C. Bennett and C. E. Wieman, Phys. Rev. Lett. **83** (1999) 889.

[30] M.-A. Bouchiat and J.Guéna, J. Phys. (France) **49** (1988) 2044.

[31] J. Guéna, M. Lintz and M.-A. Bouchiat, J. Opt. Soc. Am. **B22** (2005) 21.

---

[32] A similar extension has been implemented in the longitudinal field configuration, see Fig. 5 in Ref. [4].

[33] D. Sarkisyan, T. Varzhapetyan, A. Sarkisyan, Y. Malakyan, A. Papoyan, A. Lezama, D. Bloch and M. Ducloy, Phys. Rev. **A 69** (2004) 065802 and references therein.